# Calculation of the specific heat in ultra-thin free-standing silicon membranes

E. Chávez*[1, 2], J.Cuffe[1], F. Alzina[1], C. M. Sotomayor Torres[1, 2, 3]

[1] Catalan Institute of Nanotechnology (CIN2-CSIC), Campus de la UAB 08193 Bellaterra (Barcelona), Spain.

[2] Dept. of Physics, Universitat Autònoma de Barcelona, Campus de la UAB, 08193 Bellaterra (Barcelona), Spain.

[3] Catalan Institution for Research and Advanced Studies (ICREA) 08010 Barcelona, Spain.

*: E-mail: **emigdio.chavez@icn.cat**

**Abstract**. The specific heat of ultra-thin free-standing membranes is calculated using the elastic continuum model. We first obtain the dispersion relations of the discrete set of acoustic modes in the system. The specific heat is then calculated by summing over the discrete out-of-plane wavevector component and integrating over the continuous in-plane wavevector of these waves. In the low-temperature regime ($T < 4$ K), the flexural polarization is seen to have the highest contribution to the total specific heat. This leads to a linear dependence with temperature, resulting in a larger specific heat for the membrane compared to that of the bulk counterpart.

## 1. Introduction

The quest for enhanced thermoelectric properties generally requires large values of the figure of merit $ZT$ [1]. Recent experimental and theoretical reports point to an enhancement of the figure of merit in thin films [2], nanowires [3] and superlattices [4]. This increase of $ZT$ is attributed to the decrease of the thermal conductivity, which is predicted to be related in part to the modification of the acoustic dispersion relation due to periodicity (superlattices) or spatial confinement of the phonon modes (thin films or nanowires). This gives rise to a wavevector dependence of the group velocity and a change in the density of states [5], and a corresponding increase of the relaxations rates [6]. Moreover, any change in the acoustic dispersion relation also affects the specific heat capacity. The specific heat in an $n$-dimensional system modeled with a Debye-like dispersion relation in the ultra-low temperature regime is approximately proportional to $T^n$. Experimental measurements of the specific heat in single-walled carbon nanotubes have shown that the temperature dependence shifts from being proportional to $T^3$ (3D-like behavior) to being proportional to $T$ (1D-like behavior) [7]. However, theoretical work has also predicted similar "1D behavior" in graphene [7] due to the modification of the dispersion relation and, specifically, to the predominant contribution of the fundamental flexural mode.

In this paper we present an analysis of the effects of confinement of acoustic phonons in the computation of the specific heat of free-standing silicon membranes. The thickness and temperature dependencies are also presented.

**2. Theory**

2.1. Dispersion relation

The acoustic dispersion relation in a free-standing structure was calculated by Lamb [8] using the elastic continuum model. This model provides an adequate description of the elastic waves at long wave lengths, including nanostructures where confinement effects are observed.

The waves supported by this membrane are solutions of the elasticity equation of material with stress free at the boundaries in $z = \pm a/2$, where $a$ is the thickness of the membrane, with infinite extent in $x$ and $y$ directions. For free-standing membranes the normal components of the stress tensor vanish on the surface. The acoustic equation of motion is given by

$$\frac{\partial^2 \vec{U}}{\partial t^2} = S_T^2 \nabla^2 \vec{U} + (S_L^2 - S_T^2) \vec{\nabla}(\vec{\nabla} \cdot \vec{U}) \qquad (1)$$

where $\vec{U}$ is the displacement vector, and $S_L$ and $S_T$ are the longitudinal and transverse sound velocities respectively. This system has two types of solutions; solutions with displacements confined to the sagittal $(x, z)$ plane are called Lamb waves, while solutions with displacements perpendicular to the sagittal plane are called shear waves (SW). The Lamb waves can be further divided into two categories of modes. Those with out-of-plane symmetric and antisymmetric displacements with respect to middle plane of the plate are known as dilatational waves (DW) and flexural waves (FW), respectively [8]. The dispersion relation for shear waves has the following analytical expression [9]

$$\omega_n = S_T (q_{z,n}^2 + q_{//}^2)^{1/2} \qquad (2)$$

where $q_{z,n} = n\pi/a$ takes only a discrete set of values at each in-plane wavevector $q_{//}$ since $n$ is an integer. In contrast, the dispersion relation of saggital waves cannot be written in a simple analytical form. Instead, it must be found solving the dynamical matrix, which yields the expression

$$\frac{4 q_{//}^2 q_{l,n} q_{t,n}}{(q_{t,n}^2 - q_{//}^2)^2} = -\left( \frac{\tan(q_{t,n} a/2)}{\tan(q_{l,n} a/2)} \right)^{\pm 1} \qquad (3)$$

where the exponents $+1$ and $-1$ correspond to symmetric and antisymmetric modes, respectively. The parameters $q_l$ and $q_t$ represent the longitudinal and transverse perpendicular wavevectors. The dispersion relation is then found through the relationship between the two perpendicular wavevectors

$$\omega_n^2 = S_L^2 \left( q_{//}^2 + q_{l,n}^2 \right) = S_T^2 \left( q_{//}^2 + q_{t,n}^2 \right) \qquad (4)$$

By introducing Eq. (4) in Eq. (3) a non-linear equation is obtained, where for each value of $q_{//}$ there are many values for $q_{t,n}$ and $q_{l,n}$. The total number of modes, $n$, in the membrane is limited by the number of atoms in the thickness $a$. Thus, for a membrane with N monolayers in the thickness $a$, $n$ takes any integer value between 0 to 3N [9].

2.2. Specific heat.

The specific heat is defined as the amount of energy per unit mass or per unit volume to be supplied to a system to increase its temperature by one degree Kelvin. It can be defined as the temperature derivate of the average energy ($E$) [11].

$$C_V = \frac{1}{V}\frac{\partial E}{\partial T}\bigg|_V = \frac{1}{V}\sum_{q,p}\hbar\omega_{qp}\frac{\partial n_{qp}}{\partial T} \qquad (5)$$

where $C_V$ is the specific heat, $V$ is the volume and $n_{qp}$ is the Bose-Einstein equilibrium phonon distribution function. In the case of the membrane, where the perpendicular component of the wavevector is discretized, the dispersion relation is not Debye-like and so we cannot integrate over all $q$-space. Therefore, we keep the summation on the perpendicular component over the discrete number of modes $n$ and rewrite Eq. (5) as:

$$\begin{aligned}C_V &= \frac{1}{Sa}\sum_{n,p}\int_0^{q\max}\frac{(\hbar\omega_{qp})^2}{KT^2}n_{qp}(n_{qp}+1)\frac{S}{(2\pi)^2}(2\pi)q_{//}dq_{//} \\ &= \frac{1}{2\pi a}\sum_{n,p}\int_0^{q\max}c(\omega_{qp})q_{//}dq_{//}; \quad c(\omega_{qp})\equiv\frac{1}{K_B}\left(\frac{\hbar\omega_{qp}}{T}\right)^2 n_{qp}(n_{qp}+1)\end{aligned} \qquad (5)$$

where $c(\omega_{qp})$ is defined as the spectral density of the specific heat capacity which represents the contribution of the states in the interval $\omega + d\omega$ to the specific heat, $S$ is the total area of the membrane and $q_{max}$ is the maximum value of the parallel wavevector limited by the Debye cutoff [12].

**3. Results and discussion.**

The numerical analysis of dispersion relation and specific heat was calculated using a routine developed with the commercial software MATLAB$^{TM}$. The material parameters used in the simulation are shown in Table 1.

**Table 1**. Silicon parameters at room temperature used in the calculations

| Parameter | Symbol | Value |
|---|---|---|
| Debye radius[a] | $q_D$ | $1.140\times10^{10}$ m$^{-1}$ |
| Lattice constant[b] | $a_0$ | 0.543 nm |
| Longitudinal velocity[c] | $S_L$ | 8.433 Kms$^{-1}$ |
| Transversal velocity[c] | $S_T$ | 5.844 Kms$^{-1}$ |

[a] Huang et al. [12]
[b] R. Kahn [13]
[c] McSkimin et al. [14]

3.1. Phonon spectrum

Figure 1 shows the SW acoustic dispersion relation of a 10 nm thick membrane. Figure 2 shows the DW and FW acoustic dispersion relation of the 10 nm thick membrane. Two interesting zones in the acoustics dispersion relation are the zero-order of DW and FW as their behavior can be described quasi-analytically. For these zones it is possible to approximate the dispersion relation linearly (zero-order DW) or quadratically (zero-order FW). In the low temperature regime, where the zero-order DW and FW modes are the most populated states, their dispersion relation are of much relevance in understanding the evolution of the specific heat capacity.

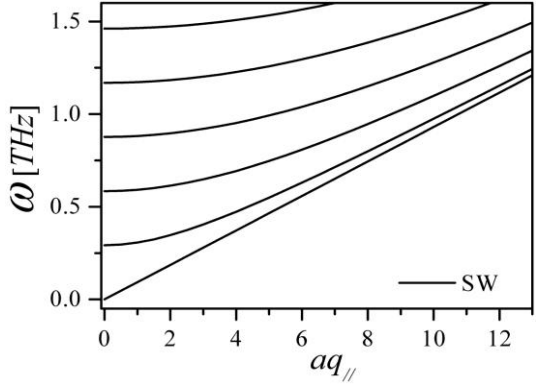 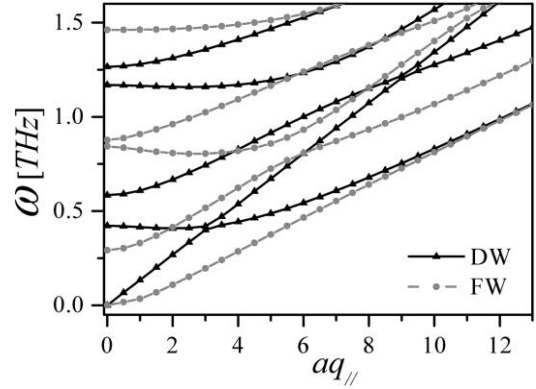

**Figure 1.** Acoustic dispersion relation of 10 nm Si membrane for shear waves (SW) as a function of the dimensionless parallel wavevector ($aq_{//}$).

**Figure 2.** Acoustic dispersion relation of 10 nm Si membrane as a function of dimensionless parallel wavevector ($aq_{//}$). The dilatational waves (DW) and the flexural waves (FW), are denoted by black triangles and solid lines and by grey dots and solid lines, respectively.

3.2. Specific heat

Once the phonon spectrum is obtained, the specific heat is calculated from Eq. (5). We take into account the discretization of the parallel component, i.e. number of branches, as well as the contribution of each polarization (SW, DW and FW) to the total specific heat.

Figure 3 shows that the temperature dependence of the specific heat for 5 and 10 nm thick Si membranes. For comparison, the specific heat of the bulk Si is also plotted. At low temperature ($T < 4$ K), the departure from a $\sim T^3$ dependence is evident, and approaches $\sim T$ as the membrane thickness decreases. This dependence reflects the predominance of the fundamental (zero-order) FW owing to its quadratic dispersion in this low temperature regime. Figure 4 shows the relative contribution of each polarization to the total specific heat for a 10 nm Si membrane. The contribution of shear waves becomes the most important above 4 K, with 38% of the total specific heat, increasing to a maximum of 43% at 30 K. Above $T > 400$ K we find a convergence of the contribution of all polarizations. Figure 5 shows that the specific heat of a 5 and 10 nm membrane becomes larger than the bulk value at a temperature threshold of approximately 40 - 50 K.

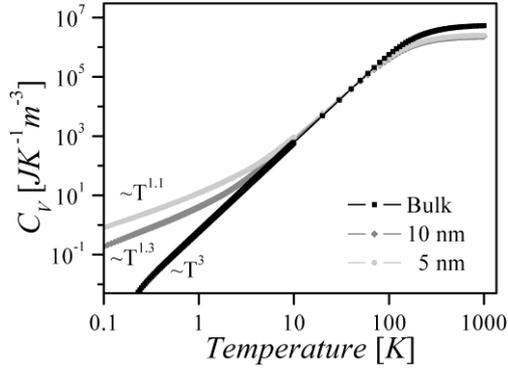
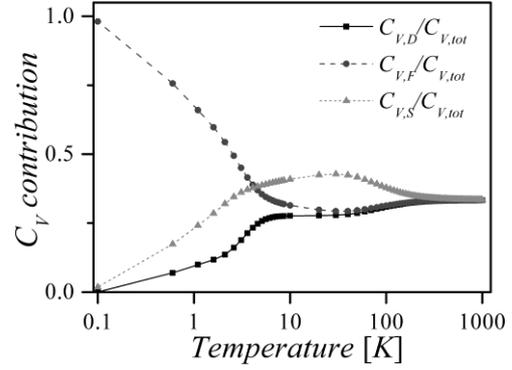

**Figure 3.** Specific heat of Si as a function of temperature for the bulk (black line), 10 nm (dark grey line) and for a 5 nm (grey line) thick membrane.

**Figure 4.** Contribution of each polarization to the total specific heat for 10 nm thick membrane. The solid black, dashed dark grey and grey dotted lines represent the contribution of dilatational ($C_{V,D}$), flexural ($C_{V,F}$) and shear ($C_{V,S}$) waves, respectively.

Figure 6 shows the specific heat as a function of the thickness. Note that while at room temperature the specific heat remains constant, at low temperature it increases with decreasing thickness. Again, this dependence reflects the dominant role of the FW in the low temperature regime.

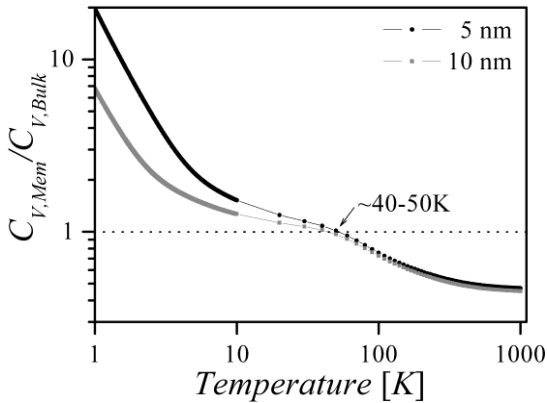
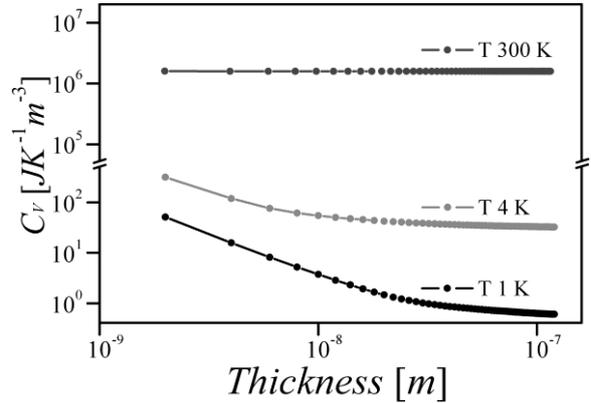

**Figure 5** Normalized specific heat for 5 (black line) and 10 nm (grey line) thick membrane. The dotted line illustrates the temperature range over which the specific heat of the bulk and the membrane are equal.

**Figure 6** Specific heat as a function of the thickness for three temperatures 300, 4 and 1 K

Figures 7 and 8 show the spectral density of the heat capacity as a function of frequency for a 10 nm thick Si membrane. The numerous peaks are due to regions of high density of states in the modified dispersion relation. It is seen that at room temperature the main contribution to the specific heat comes from of the more energetic phonons (from about 1.6 to 6 THz), while at 30 K temperature the main contribution comes from low-energy phonons (from approximately 0.3 to 1.3 THz).

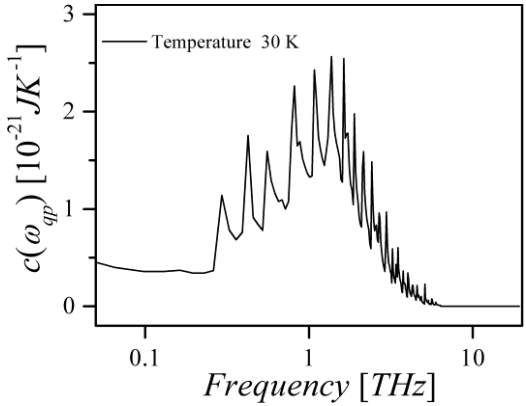

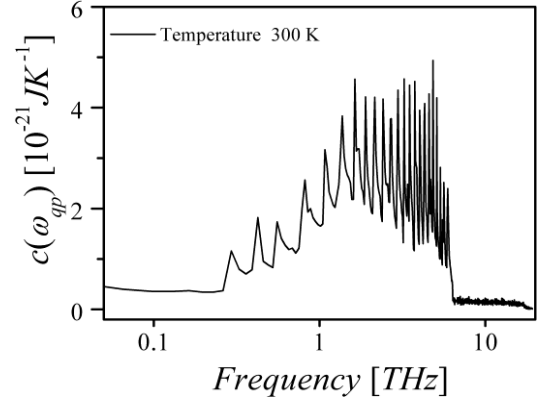

**Figure 7** Spectral density of the heat capacity of 10 nm Si membrane at 30 K as a function of frequency.

**Figure 8** Spectral density of the heat capacity of 10 nm Si membrane at 300 K as a function of frequency.

3.2.1. *Specific heat in layered systems.* Using the solutions proposed by Donnetti et al. [15], we show that is possible to use the same formulation to calculate the specific heat in layered systems. As an example, we calculated the shear acoustic dispersion relation of a 5-layer free-standing system, composed of silicon and silicon oxide with a total thickness of 3.14 nm as shown in Fig. 9. The in-plane specific heat associated to this multilayer system was compared to a 3.14 nm thick free-standing silicon membrane, silicon dioxide membrane and a membrane of an effective material formed from a weighted average of both constituent materials (Fig. 11).

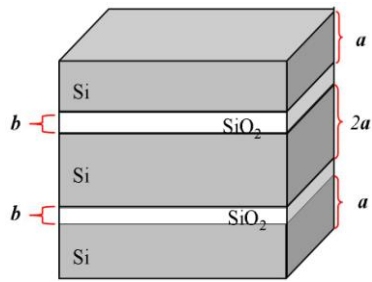
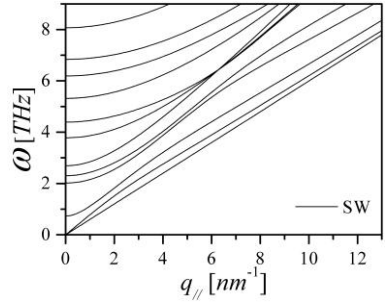
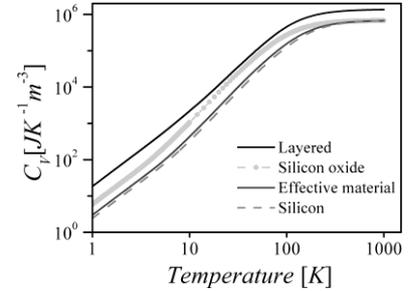

**Figure 9.** Scheme of 5-layers system, with $a = 0.54$ nm and $b = 0.49$ nm.

**Figure 10.** Acoustic dispersion relation for shear waves in the layer structure shown in figure 9.

**Figure 11.** Shear contribution to the specific heat of for the layered structure (black line), for silicon dioxide (dotted dark grey line), for an effective material (grey line) and for silicon (dashed line).

As in the case of the homogeneous membrane, the specific heat depends on the modification of the dispersion relation, which will in turn depend on the nature of the constituent materials and their thicknesses. Note that specific heat of the layered system is larger than any value of the homogeneous single-membranes, due to the appearance of localized modes within the layers, demonstrated, for example, by the splitting of the fundamental shear branch. This demonstrates the importance considering the modified dispersion relation of the system. For a complete description of the specific heat it would be necessary to include the contributions of the sagittal waves.

## 4. Conclusions

We have found that the temperature dependence of the specific heat in the low-temperature regime departs from a $T^3$ to a $T$ dependence in nanoscale free-standing silicon membranes. The change in behavior is related to the large contribution from the fundamental flexural wave, which presents a quadratic dispersion relation for small values of the in-plane wavevector. We have found that in the low temperature regime the specific heat of membranes is larger than the bulk one. For a 10 nm thick silicon membrane, the contribution of shear waves to the total specific heat becomes the most important one above 4 K, contributing over 38% of the total specific heat, increasing to a maximum of 43% at 30 K for 10 nm thick silicon membrane. Above $T > 400$ K we have found a convergence of the contribution of all polarizations. We have also shown that this theoretical model can also be used for thin multi-layered systems.

**Acknowledgments**
The authors acknowledge the financial support from the FP7 projects NANOFUNCTION (grant nr. 257375) and NANOPOWER (grant nr. 256959); the Spanish MICINN projects nanoTHERM (grant nr. CSD2010-0044) and ACPHIN (FIS2009-10150). E.C. gratefully acknowledges a Becas Chile 2010 CONICYT fellowship from Chilean government. J.C. gratefully acknowledges a doctoral scholarship from the Irish Research Council for Science, Engineering and Technology (ICRSET).